\documentstyle[epsfig]{aipproc}

\newcommand{\ba}{\begin{eqnarray}}
\newcommand{\ea}{\end{eqnarray}}

\begin{document}

\title{How random are random nuclei?\\
Shapes, triangles and kites}
\author{R. Bijker$^{1}$ and A. Frank$^{1,2}$}
\address{
$^{1)}$ ICN-UNAM, AP 70-543, 04510 M\'exico D.F., M\'exico\\
$^{2)}$ CCF-UNAM, AP 139-B, 62251 Cuernavaca, Morelos, M\'exico}
\maketitle

\

The energy systematics of medium and heavy even-even nuclei shows very 
regular features, such as for example the tripartite classification of 
nuclear structure into seniority, anharmonic vibrator and rotor regions 
\cite{Casten,Zamfir}. Traditionally, this regular behavior has been 
interpreted as a consequence of particular interactions, such as an 
attractive pairing force in semimagic nuclei and an attractive 
neutron-proton quadrupole-quadrupole interaction for deformed nuclei. 

It came as a surprise, therefore, that recent studies of even-even nuclei 
in the nuclear shell model \cite{JBD,BFP,JBDT} and in the interacting boson 
model (IBM) \cite{BF1,BF2,KZC} with random interactions also displayed a 
high degree of order. Both models showed a statistical preference 
for $L=0$ ground states, despite the random nature of 
the interactions. In addition, in the shell model evidence was found 
for the occurrence of pairing properties \cite{JBDT}, and in the IBM for 
vibrational and rotational bands \cite{BF1,BF2}. 

These unexpected results have sparked a large number of investigations 
to try to understand their origin \cite{BF4}. In 
this contribution, we discuss the phenomenom of emerging regular spectral 
features from the IBM with random interactions, and its 
relation with the underlying geometric shapes and critical points. 

\

\begin{figure}
\centerline{\epsfig{file=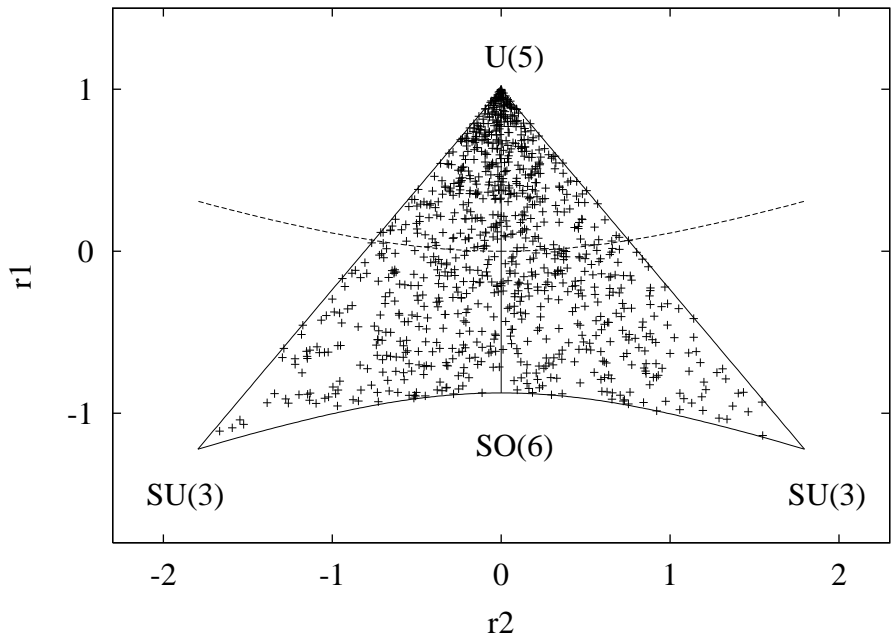}}
\caption[]{Shape phase diagram for the random CQF Hamiltonian obtained 
for $N=16$ and 1000 runs. The dashed line separates the spherical 
from the deformed shape.}
\label{r1r2}
\end{figure}

In order to study the geometric shapes associated with the IBM 
\cite{DSI,IBM,IZC}, we first consider the schematic Hamiltonian of 
the consistent-Q formulation (CQF) \cite{CQF} 
\ba
H \;=\; \epsilon \, n_d - \kappa \, Q(\chi) \cdot Q(\chi) ~. 
\nonumber
\ea
The parameters are restricted to the `physically' allowed region, 
i.e. $\epsilon>0$, $\kappa>0$ and $-\sqrt{7}/2 \leq \chi \leq \sqrt{7}/2$. 
The properties of the CQF Hamiltonian are investigated by 
taking the scaled parameters $\eta=\epsilon/[\epsilon+4\kappa(N-1)]$ and 
$\bar{\chi}=2\chi/\sqrt{7}$ randomly on the intervals $0 \leq \eta \leq 1$ 
and $-1 \leq \bar{\chi} \leq 1$. In Fig~\ref{r1r2} 
we present the results in a shape phase diagram as a function of the 
coefficients $r_1$ and $r_2$ which were introduced in \cite{ELM} as the 
essential control parameters to classify the equilibrium configurations 
of the IBM Hamiltonian. $r_1$ and $r_2$ are determined by particular 
combinations of the interaction parameters. 
In Fig.~\ref{r1r2} we have identified each of the dynamical symmetries 
of the IBM: $U(5)$, $SU(3)$ with prolate/oblate symmetry and $SO(6)$. 
The transitions between the dynamical symmetries are indicated by the 
solid lines. The resulting figure is that of a kite. The socalled critical 
point symmetries $E(5)$ and $X(5)$ \cite{FI} are related to the points 
at the intersections of the solid lines and the separatrix 
(dashed line) which separates the spherical 
and deformed shapes. The prolate and oblate deformed shapes are separated 
by $r_2=0$, $r_1<0$. The associated critical point symmetry 
coincides with the $SO(6)$ limit \cite{Jolie}. 

\begin{figure}
\centerline{\epsfig{file=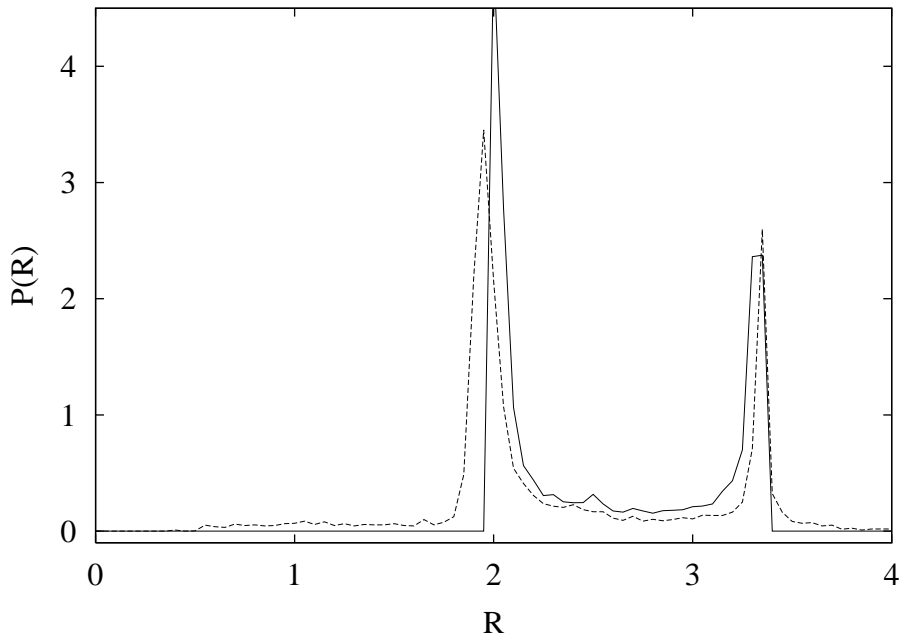}}
\caption[]{Probability distribution $P(R)$ of the energy ratio $R$ 
in the IBM for the random CQF (solid line) and for the general case 
of random one- and two-body interactions (dashed line), 
obtained for $N=16$ and 10000 runs.}
\label{pr}
\end{figure}

Each CQF Hamiltonian corresponds to a point in the $r_2 r_1$ plane and is 
labeled by a $+$ sign in Fig.~\ref{r1r2}. The random ensemble of CQF 
Hamiltonians covers the interior part of the kite: 
50$\%$ for the spherical shape, and 25$\%$ each for 
the prolate and oblate deformed shapes. For all cases, the ground state 
has angular momentum $L=0$. The existence of two definite geometric shapes, 
a spherical and an axially symmetric deformed one, is also evident from a 
plot of the probability distribution $P(R)$ of the energy ratio 
$R=[E(4_1)-E(0_1)]/[E(2_1)-E(0_1)]$. Fig.~\ref{pr} shows that for the CQF 
there are two characteristic peaks, one at the vibrator value $R=2$ and 
one at the rotor value $R=10/3$ (solid line).  

\begin{figure}
\centerline{\epsfig{file=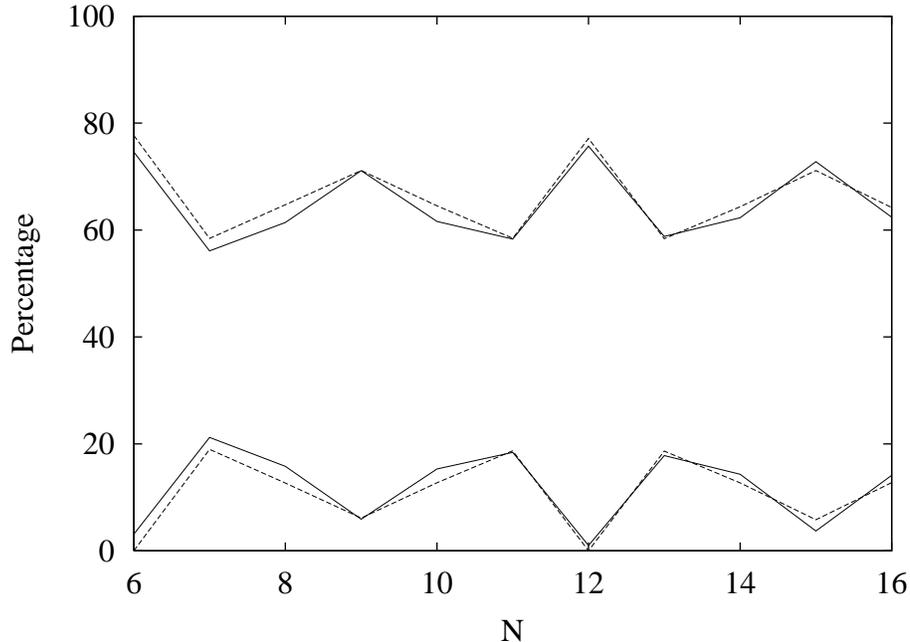}}
\caption{Percentages of ground states with $L=0$ (top) and $L=2$ (bottom) 
in the IBM with random one- and two-body interactions calculated exactly 
for 10000 runs (solid line) and in mean-field approximation (dashed line).}
\label{ibmgs}
\end{figure}

So far, we have discussed the properties of a random ensemble of IBM 
Hamiltonians with minimal constraints to `realistic' interactions.
Surprisingly enough, the results for a general IBM Hamiltonian 
with random one- and two-body interactions chosen independently from a 
Gaussian distribution of random numbers with zero mean and width 
$\sigma$ are very 
similar \cite{BF1}. Also in this case, the probability distribution 
$P(R)$ exhibits peaks at $R \sim 1.9$ and $R \sim 3.3$ (dashed line). 
The vibrational and rotational nature of these peaks has been confirmed 
by a simultaneous study of the quadrupole transitions between the 
levels \cite{BF1}. 
Despite the random nature of the interaction strengths both in relative 
size and sign, the ground state still has $L=0$ in $\sim 63\%$ of the cases. 
In Fig.~\ref{ibmgs} we show the percentages of ground states with $L=0$ 
and $L=2$ as a function of the boson number $N$ (solid line). 
We see a clear dominance of ground states with $L=0$ with $\sim$ 60-75$\%$. 
For $N=3k$ (a multiple of 3) we see an enhancement for $L=0$ 
and a decrease for $L=2$. The sum of the two hardly depends on the 
number of bosons. 

These are surprising results in the sense that, according to the 
conventional ideas in the field, the occurrence of $L=0$ ground states 
and the existence of vibrational and rotational bands are due to very 
specific forms of the interactions. 
The basic ingredients of the numerical simulations, both for the 
nuclear shell model and for the IBM, are the structure of the 
model space, the ensemble of random Hamiltonians, the order of the 
interactions (one- and two-body), and the global symmetries, i.e. 
time-reversal, hermiticity and rotation and reflection symmetry. 
The latter three symmetries cannot be modified, 
since we are studying many-body systems whose eigenstates have real 
energies and good angular momentum and parity. It has been shown 
that the observed spectral order is a robust property that does 
not depend on the specific  choice of the ensemble of 
random interactions \cite{JBD,BFP}, the time-reversal 
symmetry \cite{BFP}, or the restriction of the Hamiltonian to one- 
and two-body interactions \cite{BF2}. These results suggest that 
that an explanation of the origin of the observed regular features 
has to be sought in the many-body dynamics of the model space 
and/or the general statistical properties of random interactions. 

For the IBM, the emergence of regular features from random interactions 
can be explained in a Hartree-Bose mean-field analysis of the random 
ensemble of Hamiltonians, in which different regions of the parameter 
space are associated with particular intrinsic states, which in turn 
correspond to definite geometric shapes \cite{BF3}. There are three 
solutions: a spherical shape  
carried by a single state with $L=0$, a deformed shape which corresponds 
to a rotational band with $L=0,2,\ldots,2N$, and a condensate of quadrupole 
bosons which has a more complicated angular momentum content. We note, that 
the latter solution does not occur for the random ensemble of CQF 
Hamiltonians. The ordering of rotational energy levels depends on the sign 
of the corresponding moments of inertia, which have been evaluated with the 
Thouless-Valatin formula.  
In Fig.~\ref{ibmgs} we show the percentages of ground states with 
$L=0$ and $L=2$ as a function of the number of bosons $N$. 
A comparison of the results of the mean-field analysis (dashed lines) 
and the exact ones (solid lines) shows excellent agreement. The oscillations 
with $N$ are entirely due to the contribution of the condensate of quadrupole 
bosons. The mean-field analysis explains both the distribution of ground 
state angular momenta and the occurrence of vibrational and rotational 
bands. The same conclusions hold for the vibron model for which a large 
part of the results has been obtained analytically \cite{BF3}. 

\

In this contribution, we addressed the origin of the regular features 
obtained in numerical studies of the IBM with random interactions, 
in particular the dominance of $L=0$ ground states and the occurrence 
of vibrational and rotational band structures.  
It was shown that the geometric shapes associated with IBM Hamiltonians 
play a crucial role in understanding these regular properties. 
Different regions of the parameter space are associated with 
definite geometric shapes, such as spherical and deformed shapes and 
a condensate of quadrupole bosons. For a random ensemble of CQF 
Hamiltonians the latter solution is absent, and the shape phase diagram 
assumes the simple form of a kite or a double triangle. 

For the nuclear shell model the situation 
is less clear. Although a large number of investigations to explain and 
further explore the properties of random nuclei have shed light on various 
aspects of the original problem, i.e. the dominance of $L=0$ ground states, 
in our opinion, no definite answer is yet available, and the full 
implications for nuclear structure physics are still to be clarified.  

\

It is a great pleasure to dedicate this contribution to Rick Casten 
on the occasion of his 60th birthday in appreciation of the numerous 
occasions where his profound and stimulating comments have had a strong 
impact on our work. In particular, we gratefully acknowledge his 
enthousiastic support of the random ideas of his theoretician friends. 

\

This work was supported in part by CONACyT under projects 
32416-E and 32397-E, and by DPAGA-UNAM under project IN106400.

\end{document}